\documentclass[twocolumn,showpacs,  preprintnumbers, superscriptaddress, amssymb,pra]{revtex4-1}

\usepackage{amssymb}
\usepackage{amsmath}
\usepackage{graphicx}
\usepackage{dcolumn}
\usepackage{bm}
\usepackage{verbatim}
\usepackage{array}
\usepackage{epsfig}
\usepackage{amsbsy}
\usepackage{color}
\usepackage{soul}
\usepackage{url}
\usepackage{hyperref}
\usepackage{natbib}
\usepackage{longtable}

\usepackage{MnSymbol}

\usepackage[mediumqspace,mediumspace,amssymb,Gray]{SIunits}

\newcommand{\ds}{\displaystyle}
\newcommand{\ddsum}[1]{{\displaystyle \sum_{ #1 }}}

\definecolor{darkgreen}{rgb}{0.0, 0.6, 0.6}

\def\bra#1{\mathinner{\langle{#1}|}}
\def\ket#1{\mathinner{|{#1}\rangle}}

\begin{document}


\preprint{\hfill\parbox[b]{0.3\hsize}{ }}

\title{ Quantum interference shifts in laser spectroscopy with elliptical polarization}
 

%
\author{Pedro Amaro}
\email{pdamaro@fct.unl.pt}
\affiliation{Laborat\'orio de Instrumenta\c{c}\~ao, Engenharia Biom\'edica e F\'isica da Radia\c{c}\~ao
(LIBPhys-UNL),~Departamento de F\'isica, Faculdade~de~Ci\^{e}ncias~e~Tecnologia,~FCT,~Universidade Nova de Lisboa,~2829-516 Caparica, Portugal.}




\author{Filippo Fratini}
\affiliation{Atominstitut, Vienna University of Technology, 1020 Vienna, Austria}

\author{ Laleh Safari}
\affiliation{IST Austria, Am Campus 1, A-3400 Klosterneuburg, Austria.}


\author{Aldo Antognini}
\affiliation{Institute for Particle Physics, ETH Zurich, 8093 Zurich, Switzerland.}
\affiliation{Paul Scherrer Institute, 5232 Villigen-PSI, Switzerland.}


\author{Paul Indelicato}
\affiliation{Laboratoire Kastler Brossel, Sorbonne Universit\'e Univ. Pierre et Marie Curie-Paris 06, \'Ecole Normale
Sup\'erieure,  Coll\`{e}ge de France, CNRS, Case 74, 4 place Jussieu, F-75005 Paris, France}

\author{Randolf Pohl}
\affiliation{Max-Planck-Institute of Quantum Optics, 85748 Garching, Germany.}

\author{Jos\'e Paulo Santos}
\affiliation{Laborat\'orio de Instrumenta\c{c}\~ao, Engenharia Biom\'edica e F\'isica da Radia\c{c}\~ao
(LIBPhys-UNL),~Departamento de F\'isica, Faculdade~de~Ci\^{e}ncias~e~Tecnologia,~FCT,~Universidade Nova de Lisboa,~2829-516 Caparica, Portugal.}

\date{Received: \today  }

\begin{abstract}
We investigate the quantum interference shifts between energetically close states, where the state structure is observed by laser spectroscopy. 
%
%
We report a compact and analytical expression that models the quantum interference induced shift for any admixture of circular polarization of the incident laser and angle of observation.  An experimental scenario free of quantum interference can thus be predicted with this formula. 
 Although, this study is exemplified here for muonic deuterium, it can be applied to any other laser spectroscopy measurement  of $ns-n'p$  frequencies of a nonrelativistic atomic system, via a $ns\rightarrow n'p \rightarrow n''s $ scheme.
 

\end{abstract}


\maketitle



As it is pointed by Low \cite{low1952}, a spectral line profile can only be described by a conventional Lorentzian profile up to certain limit of accuracy. 
Beyond this limit, known as the resonant approximation, the full quantum interference between the main resonant channels and other non-resonant channels makes the spectral lines asymmetric. 
Consequently, if  Lorentzian functions are employed to fit the distorted profiles, a mismatch between the obtained centroid frequency and the actual line frequency occurs \cite{lkg1994, lsp2001,lga2009, bup2013}. 
These quantum-interference (QI) induced shifts are specific for a particular measurement and its quantification is mandatory for all high-precision spectroscopy experiments aimed for a resolution beyond the resonant approximation. 
%
A known example is the case of the $1s-2s$ transition in hydrogen, 
 which prompt many QI theoretical studies applied to various experimental methods, namely continuous-wave photon scattering \cite{lsp2001, jem2002}, two-photon excitation \cite{lss2007, lga2009} and direct two-photon frequency-comb spectroscopy \cite{mpb2014}. Other atomic  systems that have been considered include the helium fine structure and  lithium hyperfine structure, where it is shown that the negligence of the QI effects are the cause of many discrepant measurements \cite{hoh2010, mhh2012, mhh2012b, mhh2014,mhh2015, mhhb2015, scs2011,bup2013}.  
%
%
Recently, the geometric and polarization properties of the QI shifts  were investigated in laser spectroscopy, both experimentally and theoretically \cite{scs2011, bup2013}, and was found that the QI shifts vanish for a particular angle of linear polarization, so-called ``magic angle". This result has been recently applied to minimize QI shifts in  laser spectroscopy of hydrogen \cite{bmk2015}.


The aim of this article is twofold: %
First, we here continue and conclude the investigation of the QI shifts in laser spectroscopy of  muonic atoms \cite{abj2015}. 
%
We confirm that the conclusions of \cite{abj2015}, namely that the line centers are not affected on a relevant level by QI effects in muonic atoms, holds true even for an hypothetical  admixture of circular polarized light. 
Second, we extend the theoretical description developed for linear polarized photons in Ref.~\cite{bup2013}, to elliptical polarized photons. A compact and analytical formula that models the QI shifts for any  angle of observation, angle and degree of circular polarization is  presented here that forthcoming laser experiments might benefit from.

%

Laser spectroscopy is often modeled by the physical process of resonant photon scattering \cite{bup2013, bmk2015, abj2015}. 
Here, we quantify the QI effects involved in the  precise determination of the $ns-2p$ transition frequencies, by exciting the $ns\rightarrow2p$ transition, and detecting the $2p\rightarrow1s$ florescence decay. The overall process to be considered is thus $ns\rightarrow2p\rightarrow1s$ photon scattering. 
Following the second-order theory of Kramers-Heisenberg \cite{rlo2000}, the differential scattering cross section of photon scattering from an initial $2s_{J_i}^{F_i}$  state to a final $1s_{J_f}^{F_f}$  state is given in atomic units by
\begin{eqnarray}
\frac{d\sigma}{d\Omega}( \bm{\hat{\varepsilon}}_1, \theta ) =  
 \frac{1}{(2F_{i}+1)}\sum_{\tiny \begin{array}{c} m_{i},F_f,m_{f}, J_f \\ \bm{\hat{\varepsilon}_{2}} \end{array} }\left|\mathcal{M}_{i\rightarrow f}^{\bm{\hat{\varepsilon}_{1}},\bm{\hat{\varepsilon}_{2}}}\right|^{2}  
 ,\hspace{0.305 cm}
 \label{eq:diffe_sim}
\end{eqnarray} 
where $F_i$ and $F_f$ are the initial and final total angular momenta, and $m_i$ and $m_f$ the  respective projection along the quantization axis.  The second-order amplitude $\mathcal{M}_{i\rightarrow f}^{\bm{\hat{\varepsilon}_{1}},\bm{\hat{\varepsilon}_{2}} }$ involves a summation over the entire atomic spectrum \cite{saf2012b}, which in the near-resonant region comprises only the $\nu\equiv 2p_{J_\nu}^{F_\nu}$ intermediate states. In the dipole and rotating-wave  approximation, it is given by
\begin{equation}
 \ds \mathcal{M}_{i \rightarrow f}^{ \bm{\hat{\varepsilon}}_1 \bm{\hat{\varepsilon}}_2 } =  
 \ds \ddsum{F_\nu, m_\nu, J_\nu}
\frac{ \bra{f} \alpha {\bm p }\cdot \bm{\hat{\varepsilon}}_2
\ket{\nu}\bra{\nu} \alpha {\bm p }\cdot \bm{\hat{\varepsilon}}_1  \ket{i}}{\omega_{\nu i}
-\omega_1 -i \Gamma_{\nu} /2 }  ~,
\label{Mfi}
\end{equation}
with $\omega_{\nu i}$ being the $2s_{J_i}^{F_i} -  2p_{J_\nu}^{F_\nu}$ transition frequencies between $\ket{\nu}$ and $\ket{i}$, $\Gamma_\nu$ is the $2p_J^{F}$ linewidth that is assumed to be independent of the hyperfine state $\Gamma_{2p_J^{F}}\equiv\Gamma_{2p}$,  
${\bm p}$ is the linear momentum, $\bm{\hat{\varepsilon}}_1$ and $\bm{\hat{\varepsilon}}_2$ are the polarization vectors of the
incoming and scattered photon, respectively (see below), and $\alpha$ is the fine structure constant.
 Energy conservation sets $\omega_2 - \omega_1=\omega_i - \omega_f $ between the incident ($\omega_1$) and scattered ($\omega_2$) frequencies and initial ($\omega_i$) and final ($\omega_f$) atomic state frequencies.  
%
%

%
 \begin{figure}[t]
  \centering
\includegraphics[clip=true,width=1.0\columnwidth]{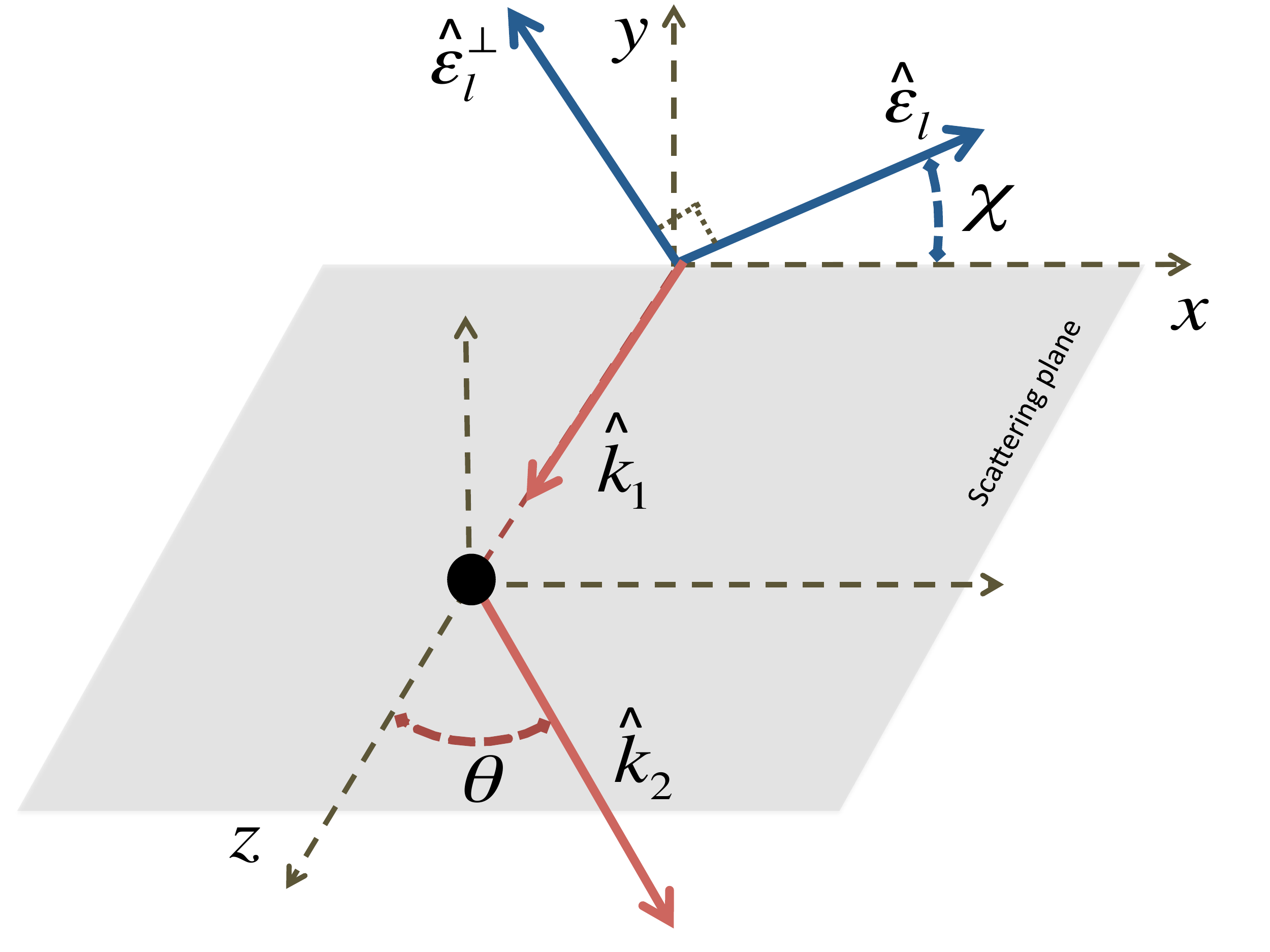}
\caption{ (Color online) Geometry of the photon scattering for incident elliptical polarized photons with momentum ${\bm k_1}$ and polarization $\bm{\hat{\varepsilon}}_1  = \left( \bm{\hat{\varepsilon}}_l  + i \eta  \bm{\hat{\varepsilon}}_l^{\perp}\right)/{\sqrt{1+\eta^2}} $, and scattered photon momentum ${\bm k_2}$, uniquely defined by $\theta$. 
$\bm{\hat{\varepsilon}_l}$ is tilted by $\chi$ relative to the scattering plane and 
$\bm{\hat{\varepsilon}^{\perp}_l}$ is orthogonal to $\bm{\hat{\varepsilon}_l}$. Both vectors are defined in the $yx$-plane. 
The second photon's polarization is not observed and is not illustrated. 
} 
\label{fig:geom}
\end{figure}
 \begin{figure*}[t]
  \centering
\includegraphics[clip=true,width=1.0\textwidth]{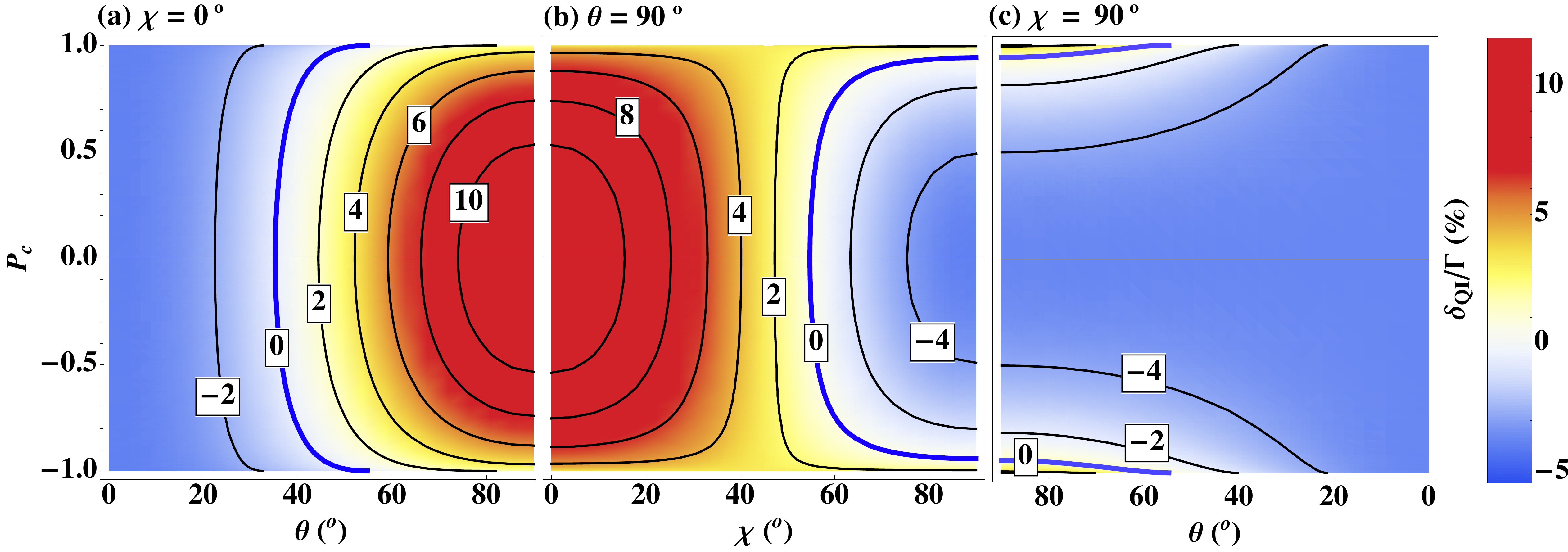} 
\caption{ (Color online) 
Contourplot of the QI shift normalized to the linewidth $\delta_{\tiny \mbox{QI}}/\Gamma$~(\%) for the $2s_{1/2}^{F=3/2} \rightarrow 2p_{3/2}^{F=1/2}$ resonance of muonic deuterium in function of the degree of circular polarization $P_c$  for the cases of (a) $\chi=0$\degree, (b) $\theta=90$\degree and (c) $\chi=90$\degree.   
The thicker blue lines show the parameter space where the QI shifts vanish.
} 
\label{fig:ellipt}
\end{figure*}

 The angular distribution of the scattered photon is given by the polar angle $\theta$, included in the scattering plane defined by both photon momenta ($\bm{k}_1$ and $\bm{k}_2$), as illustrated in Fig.~\ref{fig:geom}. 
We consider the experimental scenario of the second photon's polarization $\bm{\hat{\varepsilon}}_2$ not being detected, which is often the case in laser spectroscopy experiments \cite{ bup2013, pan2010, ans2013}. Following the procedure of Istomin {\it et al.} \cite{ipm2006}, we parametrize the incident elliptical polarized photons as $\bm{\hat{\varepsilon}}_1  = \left( \bm{\hat{\varepsilon}}_l  + i \eta  \bm{\hat{\varepsilon}}_l^{\perp}\right)/{\sqrt{1+\eta^2}} $. As shown in Fig.~\ref{fig:geom}, $\bm{\hat{\varepsilon}}_l$ is defined with an angle $\chi$ relative to the scattering plane, and  $\bm{\hat{\varepsilon}}_l^\perp= -\bm{\hat{k}}_1 \times \bm{\hat{\varepsilon}}_l$. Circular admixture is often quantified by using the degree of circular polarization $P_c$ that is defined by the difference between  left and right spherical amplitudes ($\left| \varepsilon_1^ {1} \right|^2$ and $\left| \varepsilon_1^ {-1} \right|^2$) of the incident polarization, normalized to the total amplitude ($\left| \bm{\hat{\varepsilon}}_1 \right|^2=1$). By using the previous parametrization of $\bm{\hat{\varepsilon}}_1$, it is related with the admixture parameter ($-1\le \eta\le 1$) by
%
\begin{equation}
P_c = \left| \varepsilon_1^ {1}  \right|^2-  \left| \varepsilon_1^ {-1} \right|^2    =\frac{2 \eta }{1+\eta^2}~.
\label{eq:degree_cir}
\end{equation}
By using standard angular algebra  \cite{ros1957}, Eq.~\eqref{eq:diffe_sim}  can be further rearranged in a suitable form for studying the QI shifts in terms of Lorentzian terms $\Lambda_{J_{i}J_{\nu}}^{F_{i} F_{\nu}}( \theta, \chi, \eta)$ and cross-terms $\Xi_{J_{i}J_{\nu} J_{\nu'}}^{F_{i} F_{\nu}  F_{\nu'}}( \theta, \chi, \eta)$ \cite{abj2015, bup2013}. The result is given by, 
\begin{widetext}
\begin{equation}
\frac{d\sigma}{d\Omega}( \theta, \chi,  \eta) = 
\frac{ \omega_1 \omega_2^3 \mathcal{S}_{f \nu i}^2 }{(2F_{i}+1)} 
 \left(
 \sum_{\small F_\nu,  J_\nu}  \frac{  \Lambda_{J_{i}J_{\nu}}^{F_{i} F_{\nu}}( \theta, \chi,  \eta) }{(\omega_{\nu i} - \omega_1)^2 + (\Gamma_{\nu}/2 )^2 } +   \sum_{(F'_\nu,  J'_\nu) > (F_\nu,  J_\nu)}   \frac{\Xi_{J_{i}J_{\nu} J_{\nu'}}^{F_{i} F_{\nu} F_{\nu'}}( \theta, \chi, \eta)}{    (\omega_{\nu i} - \omega_1 - i \Gamma_{\nu}/2) (\omega_{\nu'  i} - \omega_1 + i \Gamma_{\nu}/2)}
 \right) ~,
 \label{eq:Msimpli_2}
\end{equation}
\end{widetext}
with the quantities defined by  
\begin{eqnarray}
\Lambda_{J_{i}J_{\nu}}^{F_{i} F_{\nu}}( \theta, \chi, \eta) =   \sum_{m_{i},F_f,m_{f}, J_f, \bm{\hat{\varepsilon}}_{2}  }\left| \Omega_{J_{i}J_{\nu}J_{f}}^{F_{i} F_{\nu}F_{f}}(  \bm{\hat{\varepsilon}}_{1},  \bm{\hat{\varepsilon}}_{2}  ) \right|^2~, \hspace{1.0 cm}   \nonumber \\
\mbox{and} \hspace{8.0 cm}  \nonumber \\
\Xi_{J_{i}J_{\nu} J_{\nu}'}^{F_{i} F_{\nu}  F_{\nu}'}( \theta, \chi, \eta)=   \hspace{5.75 cm}   \nonumber \\  
2 \mbox{Re} \left[  \sum_{ \tiny \begin{array}{c} m_{i }, F_f\\ m_{f}, J_f,  \bm{\hat{\varepsilon}}_{2}  \end{array}}  
\Omega_{J_{i}J_{\nu}J_{f}}^{F_{i}F_{\nu}F_{f}}( \bm{\hat{\varepsilon}}_{1},  \bm{\hat{\varepsilon}}_{2}  )  
 \left(\Omega_{J_{i}J'_{\nu}J_{f}}^{F_{i} F'_{\nu}F_{f}}( \bm{\hat{\varepsilon}}_{1},   \bm{\hat{\varepsilon}}_{2}  )\right)^*  \right]~,  \hspace{0.1 cm} \nonumber \\
%
%
\label{eq:Xi}
\end{eqnarray} 
having all the geometrical and polarization dependencies in terms of 
\begin{eqnarray}
\Omega_{J_{i}J_{\nu}J_{f}}^{F_{i} F_{\nu}F_{f}}(\bm{\hat{\varepsilon}}_{1},  \bm{\hat{\varepsilon}}_{2})=  [J_{\nu} F_\nu]\sqrt{\left[F_{f},F_{i},J_{f},J_{i}\right]} \nonumber \\
 \left(\begin{array}{ccc}
J_{f} & 1 & J_{\nu}\\
1/2 & 0 & -1/2\end{array}\right)
\left(\begin{array}{ccc}
J_{\nu} & 1 & J_{i}\\
1/2 & 0 & -1/2\end{array}\right)  \nonumber \\
\left\{ \begin{array}{ccc}
F_{f} & 1 & F_{\nu}\\
J_{\nu} & I & J_{f}\end{array}\right\} \left\{ \begin{array}{ccc}
F_{\nu} & 1 & F_{i}\\
J_{i} & I & J_{\nu}\end{array}\right\}  \Theta_{F_{f} F_{i}}^{F_{\nu}}~.
\label{eq:omega}
\end{eqnarray}
Here, $I$ is the nuclear spin, $\mathcal{S}_{f \nu i}$ contains all radial integrals and 
\begin{eqnarray}
\Theta_{F_{f} F_{i}}^{F_{\nu}} =  \sum_{\lambda_1, \lambda_2}   \sum_{m_{\nu}}(-1)^{\lambda_1+ \lambda_2+m_{\nu}+m_{f}+1} \varepsilon_1^ {\lambda_1} \varepsilon_2^{\lambda_2 *}  \nonumber \\ \left(\begin{array}{ccc}
F_{f} & 1 & F_{\nu}\\
-m_{f} & \lambda_{2} & m_{\nu}\end{array}\right)\left(\begin{array}{ccc}
F_{\nu} & 1 & F_{i}\\
-m_{\nu} & \lambda_{1} & m_{i}\end{array}\right) ~.
\label{eq:bthetha}
\end{eqnarray}
%

The cross terms $\Xi(\theta, \chi, \eta)$ (angular momentum quantities are omitted for shortness) contain all interference between neighboring resonances, and if they are zero, then Eq.~\eqref{eq:Msimpli_2} is reduced to a sum of  independent Lorentzian components.  
%
%
As demonstrated by Brown {\it et al.} \cite{bup2013} for the case of linear incident polarized photons ($\eta=P_c=0$), both theoretically and experimentally, these cross-terms can be parametrized as $\Xi(\theta, \chi, 0)=b_2 P_2 (\sin \theta \cos \chi)$ (angles defined in our geometry \cite{abj2015}), where $P_2(x)=(3x^2-1)/2$ is the second order Legendre polynomial. The coefficient $b_2$ depends on the angular momenta of the states participating in the transition.  Therefore, there are particular combinations of $\theta'$ and $\chi'$,  where QI effects vanish that can be obtained by solving~$\Xi(\theta',\chi',0)=0$.  For the case of $\theta'=90\degree$, the angle $\chi'=\mbox{arcos}(1/\sqrt{3})\approx54.7\degree$ is referred as ``magic angle'' in the literature \cite{bup2013}. 


In order to investigate the role of elliptical polarization on the $\delta_{\mbox{\scriptsize QI}}$ shifts, and as a continuation of previous investigation \cite{abj2015}, we choose the resonance with the largest induced shift in muonic deuterium, which is the resonance $2s_{1/2}^{F=3/2} \rightarrow 2p_{3/2}^{F=1/2}$. 
Following the procedure in Ref.~\cite{abj2015}, we evaluate the $\delta_{\mbox{\scriptsize QI}}$ by fitting a simulated spectrum of Eq.~\eqref{eq:Msimpli_2}, that would be observed by a pointlike detector with a sum of Lorentzian profiles. Figure~\ref{fig:ellipt} displays the computed $\delta_{\mbox{\scriptsize QI}}$  in units of the linewidth for all values of $P_c$, and for three cases of $\chi=0\degree$ (a), $\theta=90\degree$ (b) and  $\chi=90\degree$ (c). 

 As can be evinced  in Fig~\ref{fig:ellipt}(b),  $\delta_{\mbox{\scriptsize QI}}$ is proportional to $P_2(\cos \chi)$ for linear polarized photons ($\eta=P_c=0$),  as mentioned in Ref.~\cite{ bup2013}. Consequently, $\delta_{\mbox{\scriptsize QI}}=0$ for the angle of polarization $\chi'\approx54.7\degree$. Additionally, the points at $\chi=0$\degree ($\delta_{\tiny \mbox{QI}}^{\parallel}$) and $\chi=90\degree$ ($\delta_{\tiny \mbox{QI}}^{\perp}$)  with $P_c=0$ represents  the QI shifts listed in Ref.~\cite{abj2015} ($\delta_{\tiny \mbox{QI}}^{\parallel}=12.3$\% and $\delta_{\tiny \mbox{QI}}^{\perp}=-4.9$\%).
 
  As can be observed in Fig.~\ref{fig:ellipt}(a), for $P_c=0$ and $\chi=0\degree$, there is an additional ``magic angle" of observation  $\theta'\approx35.3\degree$ where $\delta_{\mbox{\scriptsize QI}}$ vanishes. On the other hand, for the case of $P_c=0$ and $\chi=90\degree$ represented in Fig.~\ref{fig:ellipt}(c), $\delta_{\mbox{\scriptsize QI}}$ is independent of $\theta$. This is expected since the dipole pattern of the differential cross section depends only on the angle between polarization and scattered direction, which for $\chi=90$\degree~is independent of $\theta$.

 Moreover, Fig.~\ref{fig:ellipt}(b)  shows that the contribution of circular admixture to $\delta_{\mbox{\scriptsize QI}}$ is bounded by the values  of $\delta_{\tiny \mbox{QI}}^{\parallel}$~and $\delta_{\tiny \mbox{QI}}^{\perp}$ at $P_c=0$. Thus, any possible circular admixture reduces the QI contribution relative to the linear case and a point-like detector. 
 
 For  circular polarized photons ($\eta=P_c=\pm1$), $\delta_{\mbox{\scriptsize QI}}$ is independent of  $\chi$ (see Fig.~\ref{fig:ellipt}(b)) since the differential cross section  depends only on $\chi$  through the $x$-projection of ${\bm \varepsilon_1}$ in the scattering plane, given by $(\cos^2\chi +\eta^2 \sin \chi)/(1+\eta^2)$, that for $\eta=\pm1$ is constant. The value of $\delta_{\tiny \mbox{QI}}/\Gamma \approx2.8\%$ in this setting is the same as at $\chi=45$\degree~ and any $\eta$ or $P_c$, following the same reasoning.

  The  symmetry between helicities  $\left[\delta_{\mbox{\scriptsize QI}}(\theta,\chi,P_c)=\delta_{\mbox{\scriptsize QI}}(\theta,\chi, -P_c)\right]$, displayed in Fig.~\ref{fig:ellipt}, is a consequence of not considering the scattered polarization and the final magnetic sub-level structure in the measurement scheme.

The laser system employed by the Charge Radius Experiment with Muonic Atoms (CREMA) collaboration was designed for linear polarization \cite{aab2005, asa2009}, but some small admixture of 10\% of circular polarization cannot be excluded.  
Thus,  it is worthwhile to  evaluate the QI shift with this circular admixture for the CREMA geometry setup $\delta_{\mbox{\scriptsize QI}}^*$, following similar steps as performed in Ref.~\cite{abj2015}. 
The obtained value of $\delta_{\mbox{\scriptsize QI}}^*/\Gamma=0.3$\% for $P_c=\pm0.1$ can be compared with the value of $\delta_{\mbox{\scriptsize QI}}^*/\Gamma=0.13$\% \cite{abj2015}  for $P_c=0$ (linear polarization). This shift of 0.3\% of the linewidth sets a maximum threshold of $\delta_{\mbox{\scriptsize QI}}^*$ for all resonances of the muonic atoms considered. Thus, even in the remote case of the laser having a small amount of circular polarization, QI shifts can be neglected for the present experimental resolution of to date measured muonic transitions \cite{pan2010, ans2013}.  

The scattering process considered here is of dipolar type, which is characterized by an angular dependency of the form $\frac{d \sigma}{d\Omega} \sim a(\chi,P_c) + b(\chi,P_c) \cos^2\theta$ \cite{rlo2000}. Following the formula of $P_2$, this dipole angular distribution  can always be rewritten as   $a'(\chi,P_c) + b'(\chi,P_c) P_2(\sin\theta)$. 
Thus,  the cross-terms $\Xi( \theta, \chi, \eta)$ can also be expressed as $c (\chi,P_c)+ d(\chi,P_c) P_2(\sin\theta)$. The analytical forms of $c (\chi,P_c)$ and $d(\chi,P_c) $, obtained after  evaluation of Eqs.~\eqref{eq:Xi}-\eqref{eq:bthetha}, 
can be further rearranged  in order to include the $\chi$ and $P_c$ dependencies in $P_2$. This is accomplish with the help of $P_2(a \sqrt{b})=bP_2(a) +(b-1)/2$ and with Eq.~\eqref{eq:degree_cir}.  We found, after this procedure, that the cross-terms $\Xi( \theta, \chi, \eta)$ have a compact and analytical expression for the angular and polarization properties, which is given by  
\begin{equation}
 \Xi( \theta, \chi, P_c)= b_2 P_2 \left( \sin\theta \cdot \sqrt{ \frac{  \cos(2\chi)  \cdot \sqrt{1-P_c^2}+ 1}{2}} \right)  ~.
 \label{eq:xi_gene}
\end{equation}
The coefficient $b_2$ contains the information of the angular quantum numbers involved in a particular transition. The respective values for many resonances in muonic atoms are listed in Ref.~\cite{abj2015}. 
%
%
Equation~\eqref{eq:xi_gene} models the angular and polarization dependency of $\Xi( \theta, \chi, P_c)$ for any transition $ns\rightarrow n'p\rightarrow n''s$ in an atomic system, under the premise  of nonrelativistic and dipole approximation frameworks. %
We can thus use Eq.~\eqref{eq:xi_gene} to predict regions of the ``magic values'' $(\theta', \chi', P_c')$, where $\delta_{\tiny \mbox{QI}}=0$, by solving $\Xi( \theta', \chi', P_c')$ equal to zero. This can be used to design accordingly a spectroscopy experiment insensitive to line pulling effects. For example, the blue contour with $\delta_{\tiny \mbox{QI}}=0$ in Fig.~\ref{fig:ellipt}(b), that was computed numerically, is approximately equal to $\chi'=\mbox{arcos}\left[-1/\left(3 \sqrt{1-P_c^2}\right)\right]/2$, which for $\chi'=90\degree$ gives $P_c'\approx\pm0.94$. For circular polarization $P_c=\pm 1$, a quick inspection of Eq.~\eqref{eq:xi_gene} shows that $\delta_{\tiny \mbox{QI}}=0$ occurs for the angle of observation $\theta'=54.7$\degree, as also observed in Figs.~\ref{fig:ellipt}(a) and \ref{fig:ellipt}(c).

Apart from immediate application in laser spectroscopy of atomic systems, Eq.~\eqref{eq:xi_gene} might also be applied to molecular physics and chemistry, where  line mixing occurs due to interference of neighborhood molecular states. Without further observation of the internal structure of the target, the dipole pattern of photon scattering is quite general and independent of the target being an atom or a molecule \cite{dct1984}. Essentially, the angular and polarization dependency of interference shifts included Eq.~\eqref{eq:xi_gene} might be extended to molecular techniques based on photon scattering, such as resonant x-ray emission spectroscopy (XES) \cite{lag1996, htt2010}, Raman spectroscopy \cite{lvh2011, dtc2012}, and laser spectroscopy \cite{bds1997, msd2008, ghh2011}, where interference effects or line mixing might play a significant role.  

%


In summary, we investigated the contribution of an admixture of circular polarization to the QI shift, by considering incident elliptical-polarized photons.  
%
 Calculations performed for the CREMA detector setup revealed a negligible impact of QI effects for the maximum expected admixture of circular polarization.

We presented a compact and analytical expression that models the dependency of the angular and polarization properties to the QI shift. As a generalization of a similar expression for linear polarization \cite{bup2013}, this one contains the degree of circular polarization.  Although we considered here a particular resonance of muonic deuterium, as an illustrative example, this expression can  be applied to any transition $ns\rightarrow n'p \rightarrow n''s $ in any nonrelativistic atomic system. Thus, this equation can be used to design a spectroscopy apparatus to measure $ns-n'p$ frequencies in a scheme free of quantum interference shifts by optimizing the detector geometry, the laser polarization and the laser direction.


%
\begin{acknowledgments}

\vspace{0.5 cm}

This research was supported in part by Funda\c{c}\~{a}o para a Ci\^{e}ncia e a Tecnologia (FCT), Portugal,
through the projects No. \emph{PEstOE/FIS/UI0303/2011} and \emph{PTDC/FIS/117606/2010}, financed by the European
Community Fund FEDER through the COMPETE.
P.~A. acknowledges the support of the FCT, under Contract No. \emph{SFRH/BPD/92329/2013}.  
R.~P.\ acknowledges the support from the European Research Council (ERC) through StG.\ \#\emph{279765}. 
F.~F. acknowledges support by the Austrian Science Fund (FWF) through the START grant \emph{Y 591-N16}. 
L. S. acknowledges
financial support from the People Programme (Marie Curie Actions) of the European Union's Seventh Framework Programme (\emph{FP7/2007-2013}) under REA Grant Agreement No. [\emph{291734}]. 
A.~A acknowledges the support of the Swiss National Science Foundation
Projects No. \emph{200021L\_138175} and No.  \emph{200020\_159755}.

\end{acknowledgments}


\bibliography{HF_articles}

\end{document}